\documentclass[sigconf, nonacm]{acmart}
\setcopyright{none}
\acmConference[ICSE-SEIP 2026]{ICSE-SEIP '26: Proceedings of the International Conference on Software Engineering: Software Engineering in Practice}{April 12–18, 2026}{Rio De Janeiro, Brazil}
\usepackage{graphicx} % Required for inserting images
\usepackage{subfig}
\usepackage{url}
\usepackage{xspace}
\usepackage{tcolorbox}
\usepackage{listings}

\lstdefinestyle{prompt}{
    basicstyle=\rm\small,
    breaklines=true,
    columns=fullflexible,
    frame=single,
    numbers=none,
    keepspaces=false
}
\newcommand{\gemthink}{Gemini 2.0 Flash Thinking\xspace}
\newcommand{\tool}{\texttt{XBIDetective}\xspace}
\newcommand{\basetool}{\tool{}\textsubscript{base}\xspace}
\newcommand{\thinktool}{\tool{}\textsubscript{thinking}\xspace}
\newcommand{\finetunetool}{\tool{}\textsubscript{fine-tuned}\xspace}
\title{\tool: Leveraging Vision Language Models for Identifying Cross-Browser Visual Inconsistencies}

\author{Balreet Grewal}
\affiliation{
  \institution{University of Alberta}
  \city{Edmonton}
  \country{Canada}
}
\email{balreet@ualberta.ca}

\author{Marco Castelluccio}
\affiliation{
  \institution{Mozilla Corporation}
  \city{}
  \country{}
}
\email{mcastelluccio@mozilla.com}

\author{Suhaib Mujahid}
\affiliation{
  \institution{Mozilla Corporation}
  \city{Montreal}
  \country{Canada}
}
\email{smujahid@mozilla.com}

\author{Jeff Muizelaar}
\affiliation{
  \institution{Mozilla Corporation}
  \city{}
  \country{}
}
\email{jmuizelaar@mozilla.com}

\author{James Graham}
\affiliation{
  \institution{Mozilla Corporation}
  \city{}
  \country{}
}
\email{jgraham@mozilla.com}

\author{Jan Honza Odvarko}
\affiliation{
  \institution{Independent Researcher}
  \city{}
  \country{}
}
\email{odvarko@gmail.com}

\author{Cor-Paul Bezemer}
\affiliation{
  \institution{University of Alberta}
  \city{Edmonton}
  \country{Canada}
}
\email{bezemer@ualberta.ca}

\makeatletter
\newif\if@acm@printedauthorblock
\@acm@printedauthorblockfalse

\def\@typeset@author@bx{%
  \if@acm@printedauthorblock
    \global\setbox\author@bx=\vtop{}%
    \box\author@bx\hspace{\author@bx@sep}%
    \gdef\@currentauthors{}%
    \gdef\@currentaffiliation{}%
  \else
    \@acm@printedauthorblocktrue
    \bgroup
      \global\setbox\author@bx=\vtop{%
        \hsize=\textwidth   % full page width (instead of acmart's narrower box)
        \centering
        \@authorfont
        Balreet Grewal$^{1}$,
        James Graham$^{2}$,
        Jeff Muizelaar$^{2}$,
        Jan Honza Odvarko$^{3}$\footnote{}\authornote{This author was affiliated with Mozilla Corporation at the time of this work.},
        Suhaib Mujahid$^{2}$,
        Marco Castelluccio$^{2}$,
        Cor\mbox{-}Paul Bezemer$^{1}$\par\medskip

        \@affiliationfont
        \small
        $^{1}$University of Alberta \quad
        $^{2}$Mozilla Corporation \quad
        $^{3}$Independent Researcher\par\medskip

        $^{1}$\{balreet, bezemer\}@ualberta.ca \quad
        $^{2}$\{jgraham, jmuizelaar, smujahid, mcastelluccio\}@mozilla.com \quad
        $^{3}$odvarko@gmail.com
      }%
    \egroup
    \box\author@bx\hspace{\author@bx@sep}%
    \gdef\@currentauthors{}%
    \gdef\@currentaffiliation{}%
  \fi
}
\makeatother

\begin{document}

\begin{abstract}
    Browser rendering bugs can be challenging to detect for browser developers, as they may be triggered by very specific conditions that are exhibited on only a very small subset of websites. Cross-browser inconsistencies (XBIs), variations in how a website is interpreted and displayed on different browsers, can be helpful guides to detect such rendering bugs. Although visual and Document Object Model (DOM)-based analysis techniques exist for detecting XBIs, they often struggle with dynamic and interactive elements. In this study, we discuss our industry experience with using vision language models (VLMs) to identify XBIs. We present the \tool tool which automatically captures screenshots of a website in Mozilla Firefox and Google Chrome, and analyzes them with a VLM for XBIs. We evaluate \tool's performance with an off-the-shelf and a fine-tuned VLM on 1,052 websites. We show that \tool can identify cross-browser discrepancies with 79\% accuracy and detect dynamic elements and advertisements with 84\% and 85\% accuracy, respectively, when using the fine-tuned VLM. We discuss important lessons learned, and we present several potential practical use cases for \tool, including automated regression testing, large-scale monitoring of websites, and rapid triaging of XBI bug reports.

\end{abstract}

\begin{CCSXML}
<ccs2012>
   <concept>
       <concept_id>10010147.10010178.10010224.10010245.10010255</concept_id>
       <concept_desc>Computing methodologies~Matching</concept_desc>
       <concept_significance>500</concept_significance>
       </concept>
   <concept>
       <concept_id>10002951.10003260.10003300.10003302</concept_id>
       <concept_desc>Information systems~Browsers</concept_desc>
       <concept_significance>300</concept_significance>
       </concept>
   <concept>
       <concept_id>10010147.10010178.10010219.10010221</concept_id>
       <concept_desc>Computing methodologies~Intelligent agents</concept_desc>
       <concept_significance>300</concept_significance>
       </concept>
   <concept>
       <concept_id>10010147.10010178.10010179.10003352</concept_id>
       <concept_desc>Computing methodologies~Information extraction</concept_desc>
       <concept_significance>100</concept_significance>
       </concept>
 </ccs2012>
\end{CCSXML}

\ccsdesc[500]{Computing methodologies~Matching}
\ccsdesc[300]{Information systems~Browsers}
\ccsdesc[300]{Computing methodologies~Intelligent agents}
\ccsdesc[100]{Computing methodologies~Information extraction}

\keywords{AI4SE, Cross-Browser Compatibility, Web Browsers}
\maketitle

\section{Introduction}
\label{Section:intro}

Cross-browser inconsistencies (XBIs) occur when the same website renders differently across browsers due to how the browser may interpret or display a website's source code depending on the architecture of the browser~\cite{MDNCrossBrowserTesting}. Even with web standards intended to unify behaviour, differences in implementation details, feature support, or CSS and JavaScript handling remain, and these can act as signals of deeper rendering bugs in browsers. XBIs can range from subtle layout shifts to complete unavailability of a website~\cite{MDNCrossBrowserTesting} making them valuable indicators for detecting rendering bugs. For browser developers, ensuring broad website compatibility is crucial, but the process is time-consuming, especially when rendering bugs are triggered only under certain conditions. Catching these bugs before a browser update can significantly reduce post-release issues, and automated approaches offer a solution to achieve this. 

Most automated approaches for detecting XBIs involve computer vision techniques~\cite{choudhary2010webdiff, dallmeier2013webmate, saar2016browserbite} or DOM (Document Object Model) analysis~\cite{crosscheck2012, watanabe2023layout}. However, computer vision approaches face challenges with variable element detection~\cite{sabaren2018systematic}, and DOM model analysis may not capture all elements of a website such as HMTL5 \texttt{<canvas>} elements~\cite{finlay_vlmcanvasbugs}. Most XBI detection techniques are relatively dated, likely due to their limitations in handling variable or interactive elements, such as dynamic elements or advertisements, which have become increasingly common in modern websites. With recent advances in vision language models (VLMs), it may now be possible to revisit XBI detection in ways that overcome prior limitations, particularly with variable and dynamic elements, and make the techniques more practical for use on real-world websites.

This paper proposes \tool, a tool for leveraging VLMs to detect XBIs in websites rendered on different browsers. Specifically, to detect XBIs, \tool takes two screenshots of the same website in Mozilla Firefox~\footnote{\url{https://www.firefox.com}} and Google Chrome~\footnote{\url{https://www.google.com/chrome/}}, and then prompts a VLM to identify XBIs. 

We evaluated \tool on 1,052 bug reports of potential cross-browser inconsistencies, comparing its results with the ground truth from the reports. Using both an off-the-shelf thinking VLM and a fine-tuned VLM, \tool achieved precision scores of 77\% and 79\%, respectively, for identifying XBIs. Both versions of \tool also identify dynamic elements and advertisements with high accuracy. In a larger-scale analysis of 1,695 websites, \tool correctly ignored changing advertisements but struggled with dynamic elements that changed on each reload and with pop-up elements.

This study demonstrates that VLMs can meaningfully analyze visual differences in website renderings by leveraging their ability to process both visual and textual information. Thus, browser developers can leverage \tool to aide in identifying XBIs, which in turn can help to identify potential breakages before a browser update is made available to users. In summary, our main contributions are as follows:
\begin{itemize}
    \item A demonstration of how VLMs can capture XBIs by comparing website renderings.       
    \item The \tool tool to capture and identify XBIs in websites loaded in different browsers available at~\cite{artifact}. 
    \item A discussion of the lessons learned of running \tool in a large-scale analysis of 1,695 websites.
\end{itemize}

The rest of the paper further describes the study with Section~\ref{Section:rw} discussing related work. Section~\ref{Section:motivational_study} covers a motivational study. Sections~\ref{Section:tool_method} and~\ref{Section:experimental_setup} cover the methodology for \tool and the experimental setup of the study. Sections~\ref{Section:results} and~\ref{Section:lessons} present the experimental results and lessons learned. Section~\ref{Section:threats} discusses threats to validity of the study. Finally, Section~\ref{Section:conclusion} concludes the paper.
\section{Related Work}
\label{Section:rw}

\subsection{Detecting cross-browser inconsistencies}
Sabaren et al.~\cite{sabaren2018systematic} conducted a literature review on cross-browser inconsistency tools and found that most research focuses on techniques such as DOM model analysis, visual analysis, navigation model analysis, record/replay, static analysis, attribute comparison, and heuristic comparison. The authors highlight the challenges with these approaches; for instance, computer vision techniques struggle with detecting variable elements, like image carousels, and face difficulties in capturing accurate screenshots. DOM model analysis is challenged by interactive elements, different DOM models of the same webpage, and security measures that complicate DOM extraction. Navigation model analysis faces challenges with trigger state changes, unreachable states, and interactive elements. 

Many of the current tools consist of combining visual analysis with the other techniques for desktop~\cite{choudhary2010webdiff, dallmeier2013webmate, saar2016browserbite, crosscheck2012} and mobile browsers~\cite{watanabe2023layout, watanabe2019, semenenko2013browserbite}. For example, Watanabe et al.~\cite{watanabe2023layout}, building on their previous work~\cite{watanabe2019}, proposed a classification model that combines features of DOM-based analysis and computer vision techniques. Their approach, applied to mobile browsers, reports higher F1-scores for identifying external and internal layout failures.

Further, research has explored identifying the causes of XBIs~\cite{xu2018x, xfix, websee, mahajan2014root, long2021poster}. Notably, Xu et al.~\cite{xu2018x} propose X-Diag, an automated technique that aims to find the root cause of XBIs by checking if the inconsistency is caused by incompatible DOM APIs, CSS properties, or HTML elements. X-Diag achieves a precision of 89\% in identifying a root cause for an inconsistency between browsers, with a median runtime of 7.95 seconds. Currently, $\mathcal{X}$fix, a tool proposed by Mahajan et al.~\cite{xfix}, is one of the only automated techniques that generates repairs for XBIs. $\mathcal{X}$fix resolves a median of 93\% of XBIs reported by X-PERT~\cite{roy2014xpert}, an XBI detection approach. 

Some techniques have also explored detecting rendering bugs in browsers primarily using fuzzing~\cite{zhou2024janus, song2022r2z2, song2023metamong}, but it is not vastly explored. Recently, Zhou et al.~\cite{zhou2024janus} proposed J\textsc{anus}, a practical fuzzer that relies on Visual Delta Consistency, a test oracle. The intuition in the test oracle is that changes to an HTML file should be rendered either the same or differently by all browsers. J\textsc{anus} detects 31 non-crash rendering bugs, with 8 being fixed by developers. 

To the best of our knowledge, our study is the first to explore the use of visual language models as a tool for identifying content and structural XBIs. The objective of this study is to evaluate whether a tool with current state-of-the-art VLMs can replace traditional XBI detection techniques and serve as a viable tool for finding XBIs linked to underlying rendering bugs. We also evaluate \tool on real-world websites to assess its usability and effectiveness in practical scenarios.

\subsection{Visual analysis of software}

The rapid advancement of VLMs has led to their growing use for game bug detection~\cite{TaesiriCVPR2024, Taesiri_VideoGameBunny, taesiri2022largelanguagemodelspretty}. For example, Lu et al.~\cite{lu2025automatedbugframeretrieval} utilize GPT-4o to rank keyframes of a gameplay video based on how closely it matches a textual bug description. Their approach provides a method for reducing manual effort of quality assurance teams by providing an automated bug retrieval pipeline.

Similarly, VLMs have been applied to bug detection and testing of web applications~\cite{Ju_noncrasbugllm, Liu_guitest, demissie2025vlmfuzzvlm, liu2024visiondrivennoncrash, finlay_vlmcanvasbugs}. In particular, Wang et al.~\cite{wang_vlmgui} propose VETL, an end-to-end vision language model (VLM)-driven web testing technique that consists of two components: a text input generator and a target element selector. VETL effectively explores web state/action spaces and detects functional bugs, exposing issues in top-ranking commercial websites.

Further, visual analysis techniques are used in various areas of research such as regression testing~\cite{ tanno2020region, walsh_redecheck}, web page testing~\cite{finlay_canvas, althomali_webpage, stocco_web_test_repair, mahajan_htmlpresentationfailures}, and game testing~\cite{Liang_ag3, tuovenen2019mauto, paduraru_gametestingmethods, vgqa-bench}.

In game testing, Paduraru et al.~\cite{paduraru_gametestingmethods} suggest computer vision techniques for testing games. The authors point out that some current methods for aspects of game testing included using Tesseract OCR from OpenCV\footnote{\url{https://opencv.org}} for textual recognition, scene segmentation or template matching for output image recognition, and OpenPose~\cite{cao2019openpose} for animation testing. The authors further observe that automated agents for checking visual results can be effective, given that human testers are susceptible to errors. For web testing, Mahajan et al.~\cite{mahajan_htmlpresentationfailures} present a computer vision–based technique that detects and localizes presentation failures in web pages by identifying difference pixels to locate faulty HTML elements. To deal with dynamic text or images, the technique allows developers to specify those regions. Overall, the technique was able to identify 100\% of presentation failures and locate the faulty element in 93\% of cases.

For mobile applications, research has investigated the use of computer vision techniques to identify display issues in the UI of applications~\cite{Nighthawk_GUI, owl_eyes, ardito2021feature, yang2021uis, xie2020uied} and to support mobile UI testing~\cite{feng2023efficiency, liu2024make}. For example, Liu et al,~\cite{Nighthawk_GUI} propose Nighthawk, a fully automated approach to detect GUIs with display issues, and locate the region of the issue in a GUI.  

In line with these applications, we investigate the use of a VLM for identifying visual XBIs. Our approach leverages screenshots of the same website rendered in different browsers, using the VLM’s image understanding capabilities to identify potential inconsistencies. While techniques like VETL apply VLMs to support web application developers in testing the functionality of a site’s GUI, our focus is instead on assisting browser developers by detecting XBIs that may point to underlying rendering bugs. Hence, comparing our work with prior research experimentally is difficult, since the expected output of the approaches is different.

\section{Motivational Study}
\label{Section:motivational_study}

\newcommand{\vlmthink}{VLM{}\textsubscript{thinking}\xspace}
\newcommand{\vlmbase}{VLM{}\textsubscript{base}\xspace}
\newcommand{\vlmfinetuned}{VLM{}\textsubscript{fine-tuned}\xspace}

\begin{table*}[t!]
    \centering
    \begin{tabular}{l l r r r }
         \textbf{Task} & \textbf{VLM version} & \textbf{Accuracy} & \textbf{Precision} & \textbf{Recall}\\
         \midrule
         Advertisement detection & \vlmbase & 86\% & 70\% & 90\% \\
         & \vlmthink & 85\% & 72\% & 76\% \\
         & \vlmfinetuned & 85\% & 67\% & 98\% \\
         \midrule
         Dynamic element detection & \vlmbase & 83\% & 80\% & 92\% \\
         & \vlmthink & 90\% & 89\% & 93\% \\
         & \vlmfinetuned & 84\% & 83\% & 89\% \\
         \bottomrule
    \end{tabular}
    \caption{Experimental results for the base, thinking and fine-tuned versions of the VLM at detecting advertisements and dynamic elements}
    \label{tab:placeholder}
\end{table*}

We begin by investigating how well a VLM model can can effectively identify continuously changing elements such as dynamic elements and advertisements on a website page. As stated by Sabaren et al.~\cite{sabaren2018systematic}, most computer vision techniques for XBI identification struggle with variable elements such as image carousels; we aim to assess if limitations of traditional image detection methods can be overcome by leveraging a VLM.

We used screenshots captured of a list of 1,052 websites in Mozilla Firefox and Google Chrome. As described in Section~\ref{Section:tool_method} we took five screenshots of each website and overlaid them onto each other. We then prompted Gemini 2.0 Flash (\vlmbase) and \gemthink (\vlmthink) to identify advertisements and dynamic elements as shown in Listing~\ref{list:prompt_ad} and Listing~\ref{list:prompt_dyn}, respectively.

We also fine-tuned \vlmbase on a sample of 88 bug reports to create two separate models: one for detecting advertisements and one for detecting dynamic elements to assess whether the model can perform similar to the thinking model.

\begin{figure}[h!]
  \centering
\begin{lstlisting}[
    style=prompt,
    caption={Prompt template used to instruct VLM to identify advertisements.},
    label={list:prompt_ad}
  ]
Two renderings of the same website are provided, displayed in Chrome (image_1) and Firefox (image_2). Please analyze each rendering and focus on identifying any advertisements (excluding pop-ups) that might be present in either rendering. Answer the following question after doing your analysis:

1. Are there advertisement(s) in either rendering (not including pop-ups)? (Answer Yes or No) If there is, where is it?

Here is an example using the following two renderings:
<Examples 1-5>

Now, it is your turn to idenitfy advertisements in the Chrome rendering (image_1) and the Firefox rendering (image_2) as described. Please number your answer as:
1.
\end{lstlisting}
\end{figure}

We evaluated the VLMs' performance using precision, recall, and accuracy based on the model's ``Yes" or ``No" responses when identifying the presence of dynamic elements or advertisements, compared to the ground truth labelled by the first author. 

\textit{Findings:} 
\textbf{\vlmbase achieves an accuracy of 86\% at identifying advertisements in the screenshots of websites.} As shown in Table~\ref{tab:placeholder}, \vlmthink and \vlmfinetuned achieve a slightly lower accuracy of 85\%. Overall, the thinking and non-thinking, and fine-tuned VLMs perform similarly, though recall decreases for \vlmthink, and \vlmfinetuned has a notably lower precision. We also observe that \vlmthink often hallucinates, incorrectly identifying advertisements in multiple sections of a website. Additionally, we find that \vlmthink struggles to recognize advertisement placeholders such as grey boxes labelled as ``ads" that indicate an ad slot but may not contain a loaded ad. The lower precision of \vlmfinetuned reflects its higher number of false positives (97), suggesting that it frequently hallucinates the presence of advertisements Overall, all three VLM versions perform similarly at advertisement detection.

\begin{figure}[h!]
  \centering
\begin{lstlisting}[
    style=prompt,
    caption={Prompt template used to instruct VLM to identify dynamic elements.},
    label={list:prompt_dyn}
  ]
Two renderings of the same website are provided, displayed in Chrome (image_1) and Firefox (image_2). Please analyze each rendering and focus on identifying dynamic elements (sliders, carousels, progress bars, videos, dynamic graphs or charts, personalized recommendations, location-based recommendations, and real-time content) present in either rendering, excluding pop-ups. Answer the following question after doing your analysis:

1. Are there any type of dynamic element(s) (only look for sliders, carousels, progress bars, videos, dynamic graphs or charts, personalized recommendations, location-based recommendations, and real-time content) in either rendering? (Answer Yes or No) If there is, where is it? (do not include pop-ups)

Here is an example using the following two renderings:
<Examples 1-5>

Now, it is your turn to idenitfy dynamic elements in the Chrome rendering (image_1) and the Firefox rendering (image_2) as described. Please number your answer as:
1.
\end{lstlisting}
\end{figure}

\textbf{\vlmthink achieves an accuracy and recall value of 90\% and 93\% respectively at identifying dynamic elements in web renderings.} The precision achieved by the model is 89\%. As seen in Table~\ref{tab:placeholder}, \vlmbase and \vlmfinetuned achieve a lower accuracy of 83\% and 84\%, respectively, indicating that \vlmthink performs much better at identifying dynamic elements. The results from \vlmthink contain 28 false negatives from which 12 (43\%) are caused by the model misidentifying real-time based content such as a list of trending news stories. \vlmthink also does not correctly identify 4 (14\%) content carousels, 4 changing background images in websites, and 4 video players. While \vlmfinetuned does not match the performance of \vlmthink on this task, it outperforms \vlmbase, indicating that fine-tuning improves detection accuracy. Analyzing the false positives made by \vlmfinetuned, we find that the model misses 15 instances (37\%) of video players and 5 instances (12\%) of carousels. However, compared to \vlmthink, \vlmfinetuned performs better at detecting live content such as news articles, missing only 3 instances.

We find that, for the most part, \vlmthink performs better than \vlmbase and \vlmfinetuned at identifying advertisements and dynamic elements. Overall the results show that VLMs can reliably detect advertisements and dynamic elements without misclassifying them as cross-browser inconsistencies. This is encouraging, as prior work~\cite{sabaren2018systematic} has shown that computer vision techniques often struggle to correctly recognize these variable elements, leading to false positives. 

\textbf{While \vlmthink outperforms \vlmfinetuned, there are still scenarios where the fine-tuned model may be preferable.} Fine-tuning can be more cost-effective depending on the number and frequency of prompts to the VLM. Moreover, fine-tuning on a larger and more diverse dataset could mitigate the lower precision values as seen in Table~\ref{tab:placeholder} by exposing the model to a broader range of examples and cases. However, assembling sufficiently large and diverse datasets is costly and time-consuming, and poor fine-tuning practices may still bias the model toward the training data.

\begin{tcolorbox}[left=2pt, right=2pt, top=2pt, bottom=2pt]
  \textit{\textbf{Takeaway:} We find that VLMs can reliably detect advertisements and dynamic elements, addressing the limitations of earlier computer vision techniques for XBI detection. While thinking models achieve the highest performance, fine-tuned models offer comparable accuracy at lower cost, making them a practical choice for large-scale or continuous XBI monitoring.}
\end{tcolorbox}
\section{Detecting XBIs with \tool}
\label{Section:tool_method}

\begin{figure}[t!]
\centering
\includegraphics[width=0.95\columnwidth]{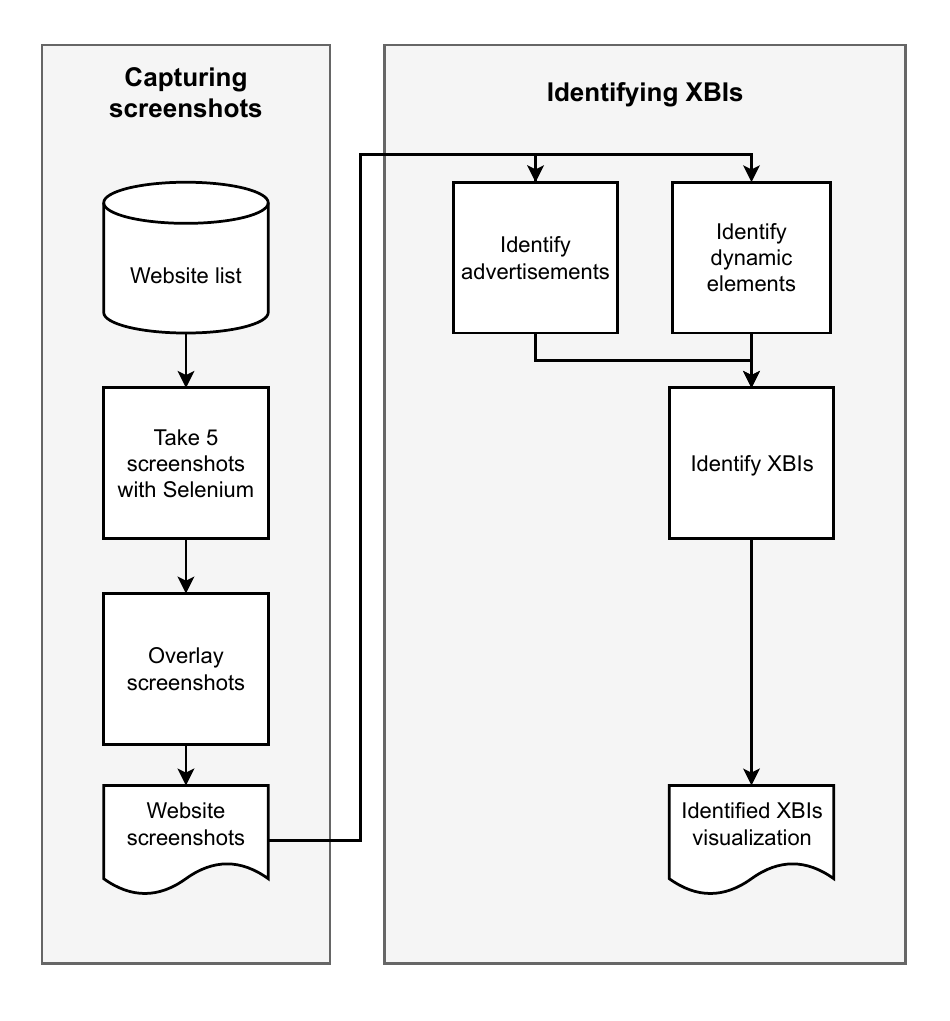} 
\caption{Overview of \tool.}
\label{fig:pipeline_overview}
\end{figure}

\begin{figure*}[t!]
\centering
\includegraphics[width=0.9\textwidth]{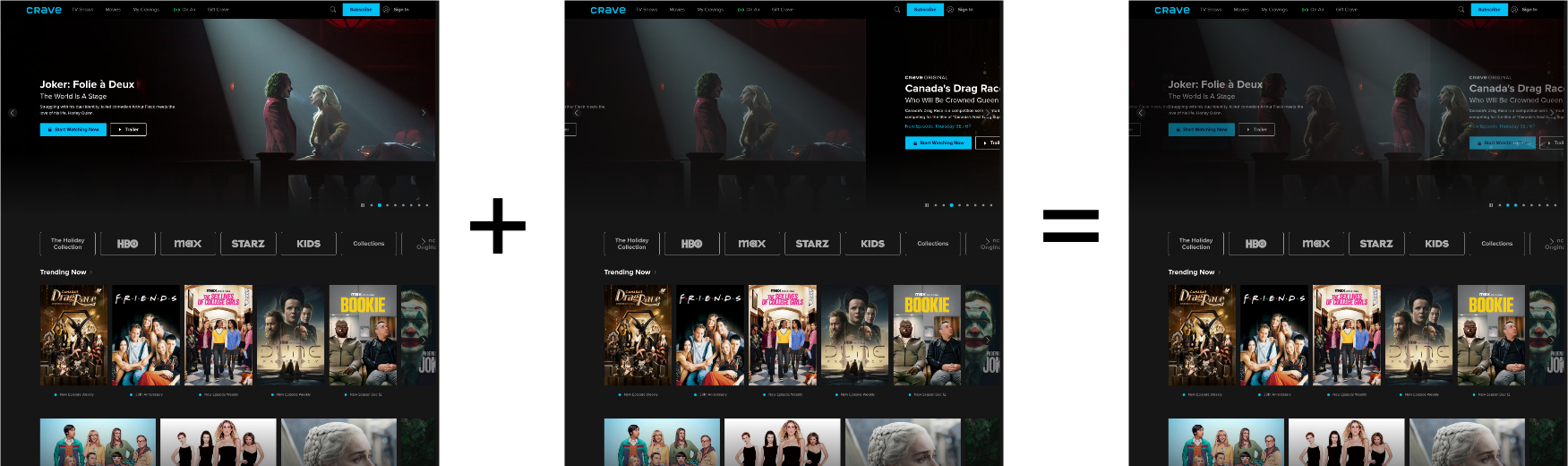} 
\caption{Example of overlay process with two screenshots taken of a website (\url{https://www.crave.ca/en}) with a dynamically changing carousel. Note that only 2 of the 5 screenshots used for the overlay are shown for brevity.}
\label{fig:overlay_ex}
\end{figure*}

\tool consists of two stages, as seen in Figure~\ref{fig:pipeline_overview}: capturing screenshots from a list of websites, and prompting the VLM for XBI identification. We explain both portions below.

\subsection{Capturing screenshots}
To capture full-page screenshots of the websites, we use Selenium\footnote{\url{https://www.selenium.dev}}, a web browser testing tool.
We capture website screenshots in two browsers running in headless mode. Five screenshots of each site are taken and overlaid to differentiate dynamic elements (e.g., image carousels) from static elements (e.g., backgrounds). For example, as shown in Figure~\ref{fig:overlay_ex}, we take screenshots at one-second intervals to capture changes in the main carousel of a website displaying video suggestions. The overlay on the right merges instances where the carousel is in transition, allowing the VLM, when prompted, to recognize the element as not static by observing the change. 

\subsection{Identifying XBIs}

A VLM that supports multiple images can be prompted to identify XBIs by providing two screenshots of the same website (cropped to the same size) rendered in different web browsers (or the same for regression testing). There are three stages to identifying XBIs.

\subsubsection{Stage 1: Advertisement detection:} In the first stage, advertisements are identified in the screenshots. Identifying advertisements ensures that we can avoid marking them as XBIs when identifying them later on. The prompt that we used is seen in Listing~\ref{list:prompt_ad}.

\subsubsection{Stage 2: Dynamic element detection:} In the second stage, the following dynamic elements are identified in the screenshots: sliders, carousels, progress bars, videos, dynamic graphs or charts, personalized recommendations, location-based recommendations, and real-time content. These elements are excluded from XBI detection because, while they may change during website rendering, such changes do not reflect inconsistencies between two websites. The VLM prompt we used is shown in Listing~\ref{list:prompt_dyn}.

\begin{figure}[t!]
  \centering
\begin{lstlisting}[
    style=prompt,
    caption={Prompt template used to instruct VLM to identify XBIs between browsers.},
    label={list:prompt_comp}
  ]
Two renderings of the same website are provided, displayed in Chrome (image_1) and Firefox (image_2). 

In terms of advertisements, this is what was found: <prompt 1 output>

In terms of dynamic elements, this is what was found: <prompt 2 ouput>

Please analyze each rendering and focus on comparing each rendering in terms of layout, font style and size, colour consistency, alignment, and the presence or absence of elements (such as buttons, text fields, advertisements, pop-ups and images). Ensure to analyze beyond a pop-up (if applicable). Ignore the advertisements and dynamic elements identified as mentioned. Answer the following questions after doing your comparison:

1. Is there any functionally or perceptually meaningful difference between the renderings (ignoring the advertisements for question 1 and ignoring the dynamic elements found in question 2 unless they cause a meaningful difference between renderings)? INCLUDING the presence or absence of pop-ups? (Answer Yes or No.)
If your answer was yes to the question 1, then answer these questions:
2. What specific element(s) are affected?
3. How do the renderings differ?
4. What would be the impact be? The impact can be evaluated by selecting the relevant label. The labels and their definitions are:
(blocked-unsupported) ... (significant-visual) ... (minor-visual) ...

Here is an example using the following two renderings:
<Examples 1-5>

Now, it is your turn to compare the Chrome rendering (image_1) and the Firefox rendering (image_2) as described. Please ensure to number your answers as:
<1.-4.>

Remember only answer questions 2-4 if your answer to 1 was yes and think about the impact while analyzing the renderings.
 
\end{lstlisting}
\end{figure}

\subsubsection{Stage 3: XBI detection:} In the final stage, XBIs are identified while ignoring the advertisements and dynamic elements detected in the previous stages. Listing~\ref{list:prompt_comp} shows the prompt we used, which includes examples to help the model understand the task. During this stage, an impact score is also assigned to each identified XBI, categorizing it into one of four severity levels. These impact scores provide Mozilla developers with guidance on which XBIs to prioritize for bug analysis. The impact scores, originally used internally at Mozilla and adapted for the task of XBI detection, are as follows:

\begin{itemize}
    \item \textbf{minor-visual}: the site has an XBI, but it does not affect the content or functionality of the site, and users are unlikely to notice. Some examples include different focus outlines on elements, small discrepancies in text rendering such as font that does not comprise readability, slight misalignments, or different background colours.
    \item \textbf{significant-visual}: the entire site does not load, the site loads but is effectively unusable, the site has visible layout problems, some parts of the page content (text, images, videos, or pop-ups) are missing or hard to access, or some features of the site are missing or broken. Some examples include, an entirely blank page, missing copy/paste buttons in a text editor, missing a pop-up on the website, or a layout that renders the website as unreadable.
    \item \textbf{blocked-unsupported}: there is a message indicating the browser is not supported. This is considered an XBI because a rendering bug might be preventing the website from displaying, even though it should be accessible to users. Additionally, the browser may be unsupported due to site requirements or rendering bugs that could potentially be addressed by the browser developer.
     \item \textbf{no-XBI}: no XBIs are observed.
\end{itemize}

Once the XBIs are identified, we generate an HTML-based visualization of the VLM's results for developers.

\section{Experimental setup}
\label{Section:experimental_setup}
\begin{figure*}[t!]
\centering
\includegraphics[width=0.95\textwidth]{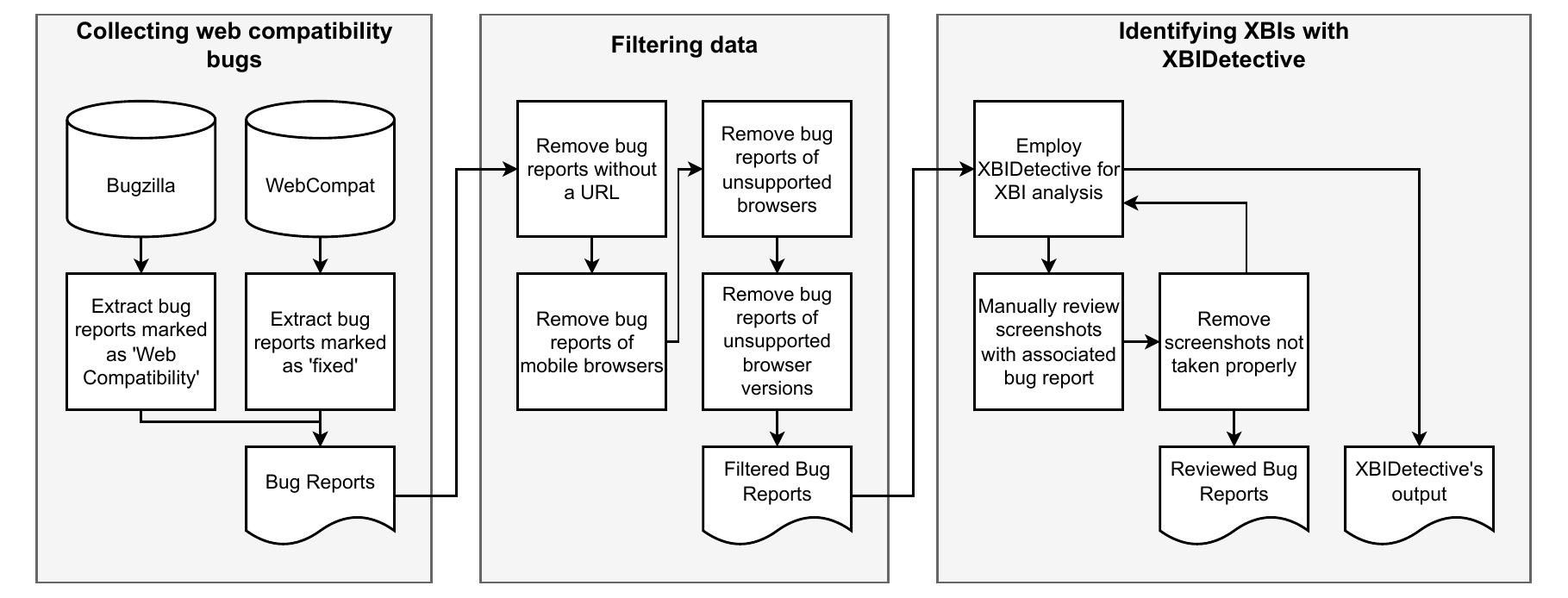} 
\caption{Overview of experimental setup.}
\label{fig:overview}
\end{figure*}

Below we describe the experimental setup, where we collected and verified web compatibility bug reports to evaluate the effectiveness of \tool at identifying XBIs. The overview of the experimental setup can be seen in Figure~\ref{fig:overview}.

\subsection{Collecting web compatibility bugs}
To conduct an evaluation of XBIs, we collected a list of websites and their corresponding bug reports of web compatibility issues from Bugzilla\footnote{\url{https://bugzilla.mozilla.org/home}}, an issue tracker for Firefox. The collected bugs are those marked as web compatibility issues in Bugzilla and that have undergone triage.
Additionally, we gathered websites and bug reports from WebCompat\footnote{\url{https://webcompat.com}}, a website dedicated to reporting web compatibility bugs. The web compatibility issues are collected from the `fixed' milestone in the WebCompat GitHub repository\footnote{\url{https://github.com/webcompat/web-bugs}}. The web compatibility tags used by both websites refer to reported XBIs for Mozilla Firefox (in both Bugzilla and WebCompat) as well as for other browsers (in WebCompat). By selecting triaged Bugzilla bugs and WebCompat bugs marked as fixed, we ensured that all collected reports corresponded to verified XBIs.

We extracted the following fields from each bug report:
\begin{itemize}
    \item \textbf{BugID}: the unique ID of the bug.
    \item \textbf{URL}: the URL of the affected website.
    \item \textbf{Browser/version}: the browser and version where the issue occurs (some reports may lack version details).
    \item \textbf{Summary}: a description of the bug.
    \item \textbf{Impact score}: If available for some Bugzilla reports, the severity of the bug as determined by the WebCompat team, used for the ground truth.
\end{itemize}

Although WebCompat allows reporters to submit screenshots, these are not included in the dataset.
In total, we collected 4,725 % 1008 + 76 + 501 + 81 + 3059
web compatibility bug reports for analysis.

\subsection{Filtering data}
We first removed bug reports that do not provide a website URL. To ensure compatibility with Selenium, bug reports collected from WebCompat were filtered to exclude mobile browsers, and all browsers except Mozilla Firefox. To maintain relevance, we also excluded reports referencing Firefox versions older than 100, given that the latest version is 143. 
After filtering, 1,052 bug reports remain. 

\subsection{Identifying XBIs with \tool} 
For this experiment, we used Selenium with Mozilla Firefox and Google Chrome. For bug reports for older Firefox versions, we reverted Selenium to that version to capture the screenshot to increase the chances of reproducing the bug.

To verify the quality of the screenshots from Selenium, the first author manually reviewed the screenshots with respect to the following criteria in order to prepare a ground truth:
\begin{itemize}
    \item Proper website loading and full-page capture. 
    \item Any XBIs found that were not originally mentioned in the bug report are appended to the report.
    \item The presence of advertisements. 
    \item The presence of the following dynamic elements: sliders, carousels, progress bars, videos, dynamic graphs or charts, personalized recommendations, location-based recommendations, and real-time content. 
    \item An impact score is assigned to each bug report if one is not already specified. The impact scoring process was calibrated in consultation with Mozilla developers. 
\end{itemize}

After manual analysis, there were screenshots from 243 websites that contained an XBI. 538 screenshots were the same across browsers, likely because the underlying XBIs had been resolved, and 271 screenshots were deemed unusable, e.g., due to being blocked by bot detectors.

We evaluated three VLMs: Gemini 2.0 Flash, \gemthink, and a fine-tuned variant of Gemini 2.0 Flash. This comparison allows us to assess the performance of a base model, a thinking model, and a fine-tuned base model in identifying XBIs. Thinking models may provide better results due to their extended reasoning capabilities, whereas base models are often more cost-effective. Finally, we analyzed whether a fine-tuned base model can achieve performance comparable to that of a thinking model.

To fine-tune Gemini 2.0 Flash, we used a statistically representative sample of 88 randomly selected bug reports (with 90\% confidence and a 10\% margin of error) from the reviewed reports described above. Supervised fine-tuning was performed using the prompt template shown in Figure \ref{list:prompt_dyn}. We omitted the bug reports used for fine-tuning during the rest of the experiments for the fine-tuned model. 

We evaluated the three VLM's performance using precision, recall, and accuracy based on the model's predicted impact scores with the impact score assigned by the first author. The metrics are defined as follows:

$$ \mathrm{Precision} = \frac{1}{N} \sum_{i=1}^{N} \frac{\mathrm{TP}_i}{\mathrm{TP}_i + \mathrm{FP}_i}$$

$$\mathrm{Accuracy} = \frac{\mathrm{TP}+\mathrm{TN}}{\mathrm{TP}+\mathrm{TN}+\mathrm{FP}+\mathrm{FN}}$$

$$\mathrm{Recall} = \frac{1}{N} \sum_{i=1}^{N} \frac{\mathrm{TP}_i}{\mathrm{TP}_i + \mathrm{FN}_i}$$
Where $N$ is the 4 impact labels.
The components of the confusion matrix are defined as follows:
\begin{itemize}
    \item \textbf{True Positive (TP):} The model assigned the correct impact score denoting the presence of an XBI to a bug report.
    \item \textbf{True Negative (TN):} The model correctly identified the bug report as not containing an XBI.
    \item \textbf{False Positive (FP):} The model identifies an XBI and assigns an impact score when there is no XBI. 
    \item \textbf{False Negative (FN):} The model identifies that a bug report has no XBI when it does contain an XBI.
\end{itemize}
The first author manually compared \gemthink’s textual output describing the identified XBIs (true positives) to the ground truth to determine the number of correctly detected XBIs.

To evaluate \tool on a broader dataset of websites, we collected a secondary dataset of 1,695 websites consisting of the top 1,000 websites with the highest number of reported bugs and 695 bug reports from WebCompat. For this dataset, we did not manually filter screenshots or establish a ground truth. We then used the fine-tuned version of \tool to analyze the screenshots and detect XBIs.

For the remainder of the paper \tool with the use of Gemini 2.0 Flash as the VLM will be referred to as \basetool, \tool with \gemthink as \thinktool, and \tool with the use of the fine-tuned version of Gemini 2.0 Flash as the VLM will be referred to as \finetunetool. 

\section{Experimental Results}
\label{Section:results}

\begin{figure}[t!]
\centering
\includegraphics[width=\columnwidth]{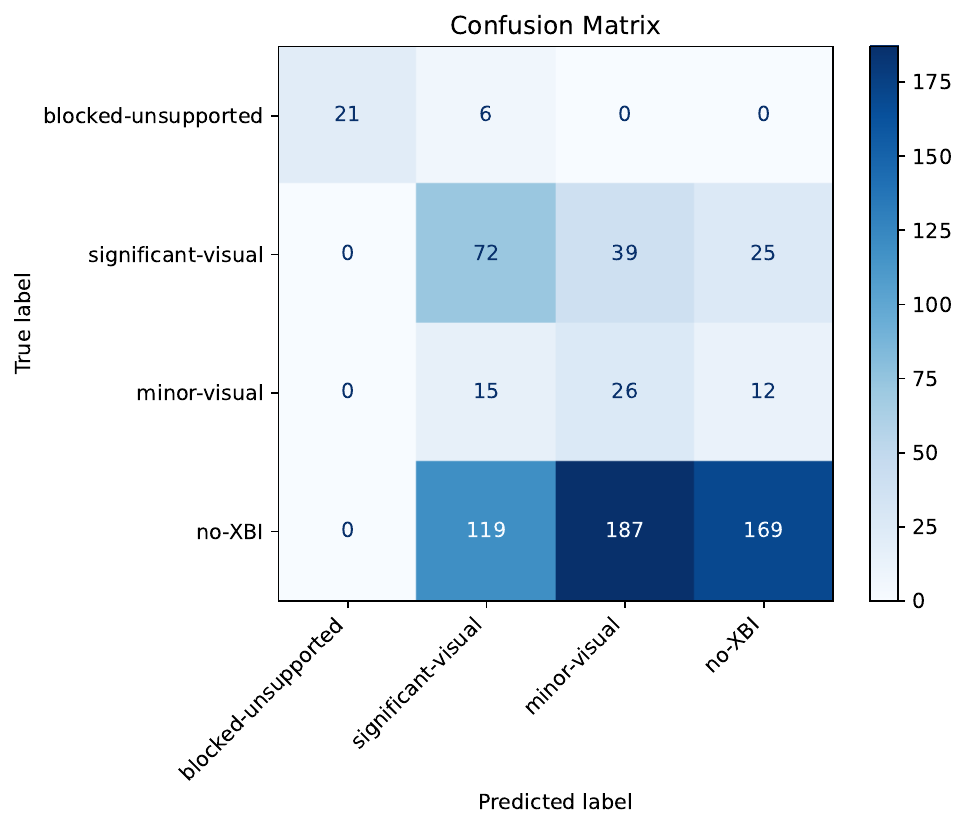} 
\caption{Confusion matrix of \basetool's labelling of the impact score.}
\label{fig:cm_base}
\end{figure}

\begin{figure}[t!]
\centering
\includegraphics[width=\columnwidth]{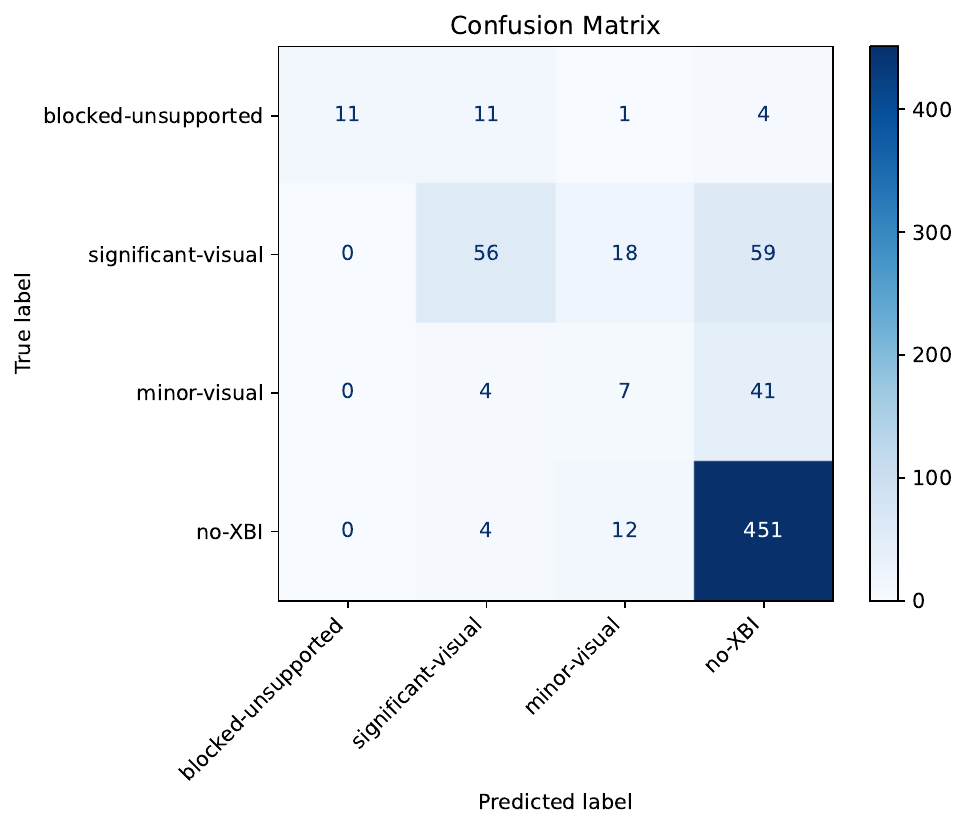} 
\caption{Confusion matrix of \thinktool's labelling of the impact score.}
\label{fig:cm}
\end{figure}

\begin{figure}[t!]
\centering
\includegraphics[width=\columnwidth]{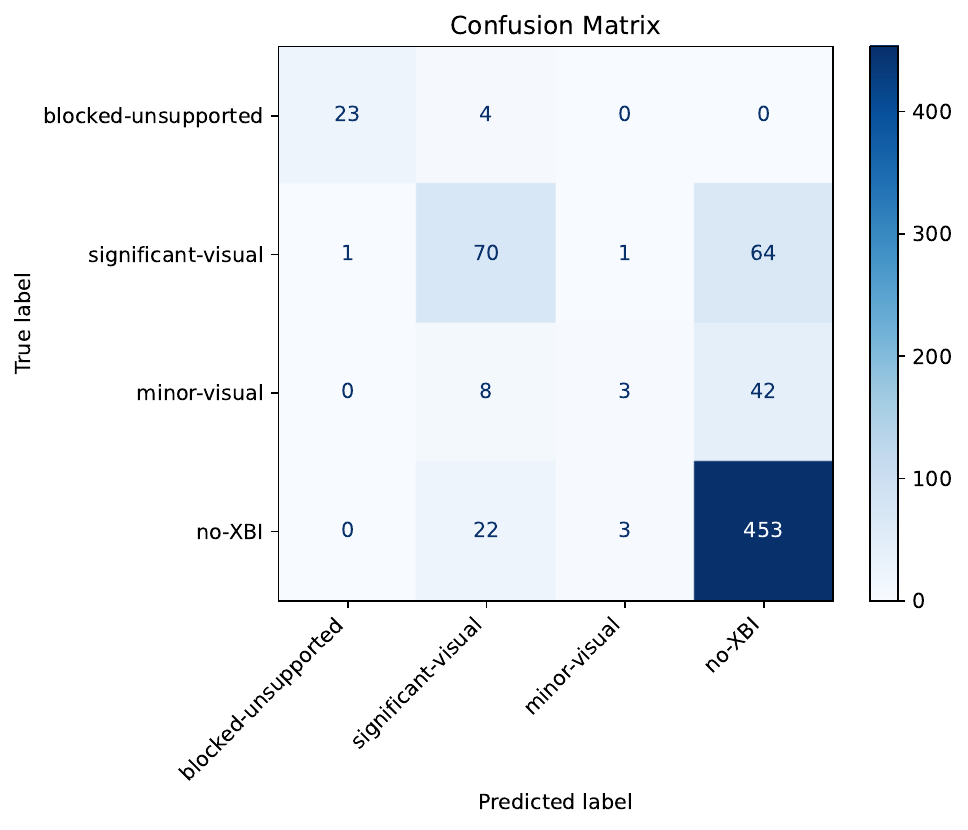} 
\caption{Confusion matrix of \finetunetool's labelling of the impact score.}
\label{fig:cm_fine}
\end{figure}

\begin{table}[t!]
    \centering
    \begin{tabular}{p{3.5cm} r r r }
         \raggedright \textbf{\tool version} & \textbf{Accuracy} & \textbf{Precision} & \textbf{Recall}\\
         \midrule
         \basetool & 42\% & 57\% & 54\% \\
         \thinktool & 77\% & 69\% & 48\% \\
         \finetunetool & 79\% & 72\% & 59\% \\
         \midrule
         \raggedright\textbf{Without advertisement detection} & & & \\
         \thinktool & 68\% & 59\% & 42\% \\
         \finetunetool & 76\% & 64\% & 58\% \\
         \midrule
         \raggedright\textbf{Without dynamic element detection} & & & \\
         \thinktool & 73\% & 63\% & 45\% \\
         \finetunetool & 76\% & 65\% & 58\% \\
         
         \bottomrule
    \end{tabular}
    \caption{Experimental results on the performance of \tool versions in assessing XBI impact score}
    \label{tab:tool_comp}
\end{table}

\textbf{\finetunetool achieves an accuracy and precision of 79\%, and 72\%, respectively at labelling the impact scores.} The recall achieved by the fine-tuned \tool is 59\%. Whereas, \thinktool achieves an accuracy and precision of 77\%, and 69\%, respectively at labelling the impact score of XBIs. The recall achieved by \thinktool is 48\%. In comparison, as shown in Table~\ref{tab:tool_comp}, \basetool performs considerably worse, achieving a precision of 42\% and producing many more incorrect classifications, as shown in Figure~\ref{fig:cm_base}. As seen from the confusion matrix in Figure~\ref{fig:cm}, \thinktool incorrectly classifies 121 web renderings as ``no-XBI" despite the presence of XBIs. Since false negatives are the most frequent type of error by the model, we examine them in more detail to understand which discrepancies the model is likely to miss. The false negatives primarily involve missed XBIs in layout differences (34 instances), the presence of pop-ups (24 instances), the dynamic elements themselves (15 instances), an image not rendering (13 instances), and the site failing to load (7 instances).

%Further, \finetunetool achieves an accuracy and precision of 79\%, and 72\%, respectively at labelling the impact scores. The recall achieved by the fine-tuned \tool is 59\%. These results are comparable to those of \thinktool, as seen in Table~\ref{tab:tool_comp}, despite not using a thinking model for the VLM, suggesting that fine-tuning a base model for \tool can perform as well as the thinking version of a VLM used in \tool. %\finetunetool also shows consistent behaviour, with an average accuracy of 79\% across five runs, and similar stability in precision and recall.

As seen in Figure~\ref{fig:cm_fine} \finetunetool incorrectly identifies 106 screenshots as ``no-XBI". These false negatives, similar to \thinktool, involve missing XBIs such as the presence of pop-ups (31 instances), font discrepancies (13 instances), and images not rendering (8 instances). Notably, 81 of the false negatives identified by the fine-tuned \tool were also identified by the base version of \tool, indicating potential ambiguity in those renderings.

Overall, \thinktool and \finetunetool perform well, indicating that they can reliably detect XBIs. Developers are mostly interested in the significant-visual and blocked-unsupported categories, as the impact of such XBIs is likely the largest. Hence if we combine the minor-visual and no-XBIs categories, we focus on the discrepancies that browser developers are most likely to prioritize.  Under this perspective, the model’s performance appears even stronger for real-world applications, achieving an accuracy of 85\% for both versions of \tool, since minor visual issues, such as a search bar with a smaller width than the page width, may not be prioritized for fixes. Nevertheless, it is important to consider the non-combined metrics, as they show how many minor issues are correctly flagged by \tool, providing a more complete picture of its detection behaviour.

\textbf{\thinktool's textual output identifying the XBI, correctly matched 92\% of XBIs labelled in the ground truth.}  In these instances, \thinktool correctly located the XBI on each page. Among the incorrect classifications (9 instances), \thinktool occasionally hallucinates issues, including misidentifying spacing inconsistencies (5 instances), and incorrectly identifying an advertisement as a website element (1 instance). Further, \tool also misclassified a dynamic element change as an XBI (1 instance). For the final 2 instances, the hallucinations made by \tool are of a change in colour between the screenshots of the websites, and the presence of a sidebar in the website.

\textbf{Explicitly directing the VLM to identify dynamic elements and ads increases the accuracy of \tool in detecting XBIs.} Without prompting \thinktool and \finetunetool to identify advertisements, their accuracy in identifying XBIs drops to 68\%, and 76\% respectively. Similarly, without prompting the \tool to identify dynamic elements, the accuracy in detecting XBIs drops from 77\% to 73\% for \thinktool and from 79\% to 76\% for \finetunetool
The drop in performance in identifying XBIs by both versions of \tool suggests that prompting the model to identify advertisements and dynamic elements,  then ignore them is an essential step within the pipeline. Without ad detection, both versions of \tool generate significantly more false positives, incorrectly classifying no-XBI sections as significant-visual (56 instances vs. 5 for \thinktool) and minor-visual (39 instances vs. 11 for \thinktool). \finetunetool's performance does not drop as drastically as \thinktool, but it is still significant enough to make the results from the pipeline less informative. Further, because dynamic elements can appear differently across browsers when a page loads, failing to identify them may lead the model to misinterpret them as XBIs, even though these variations are considered to be expected variations in a website.

\begin{tcolorbox}[left=2pt, right=2pt, top=2pt, bottom=2pt]
  \textit{\textbf{Takeaway:} Overall, \finetunetool demonstrates better performance than \thinktool in identifying XBIs.  \thinktool uses a thinking model that involves multiple reasoning steps for the VLM, making it more computationally expensive and slower to run. The cheaper, faster \basetool underperforms in identifying XBIs. However, after fine-tuning \basetool into \finetunetool, we can still use a non-thinking model which is optimized for the three specific stages to identify XBIs, allowing it to operate more efficiently in the long run.}
\end{tcolorbox}

%------------------------------------------------------
%--------------------RQ 2 -----------------------------
%------------------------------------------------------
\section{Lessons Learned}
\label{Section:lessons}
The ground truth dataset described in the previous section was carefully curated. For instance, we manually removed broken or incomplete screenshots. While this level of curation was necessary to evaluate \tool, it would not be feasible in a realistic large-scale run on many websites. To better understand how \tool performs under such conditions, we conducted a larger-scale run on 1,695 websites. From this run, \tool identified 78 XBIs (55 significant-visual, 10 minor-visual, and 13 blocked-unsupported). Note that these are realistic results, as the normal assumption is that most websites do not contain XBIs. In this section, we discuss the lessons learned from that experiment.

\subsection{Lesson 1: Capturing comparable screenshots across browsers is hard} We find that capturing consistent and comparable screenshots across different browsers presents a major challenge. The most important obstacles for capturing comparable screenshots were: 

\textbf{Obstacle 1: Rendering quirks in headless mode.} When using Selenium with Google Chrome, the browser does not support full-page screenshots unless it is run in headless mode. However, headless mode is not ideal as it can introduce its own rendering quirks~\cite{headless_medium}. Despite this limitation, we use headless mode as a necessary compromise to ensure consistent sizes across screenshots.

\textbf{Obstacle 2: Inherent stylistic differences between browsers.} Browsers behave differently and may render websites with slight variations, e.g., because of differences in how scrollbars are handled.

\textbf{Obstacle 3: Blocked websites.} Some websites blocked \tool, most likely due to anti-bot systems (such as Cloudflare). Especially Google Chrome is more susceptible to bot detection when driven by Selenium, and was frequently blocked by anti-bot systems.

The first two obstacles are difficult to overcome automatically. We found two effective ways to ignore blocked websites: 
\begin{itemize}
    \item \textbf{Preprocessing:} During the screenshot taking process, if a site displays keywords that suggest blocked access, we do not take the screenshot. Keywords include phrases such as ``403 Forbidden'' or ``you have been blocked''.
    \item \textbf{Post-inference filtering:} After the fine-tuned \tool provides its output for XBI detection, we pass the images and the \tool’s output to a secondary VLM. This model analyzes the text for any mention of a page not loading, and reviews the images to ensure they do not contain a message indicating they are blocked. Prompting the VLM allows for cases where blocked messages are in a different language or the original VLM output mentions being blocked to be filtered out, removing false positives from the report that is analyzed by developers. 
\end{itemize}

These filtering steps reduce the number of false positives that browser developers need to inspect in large-scale runs. Importantly, filtering websites has little effect on the overall usefulness of the results, since large-scale analyses are still likely to uncover multiple instances of XBIs that point to the same underlying problem.

\subsection{Lesson 2: \tool can be used for regression testing as well} In the experiments we discuss in this paper, we focused on XBIs. Another interesting application of \tool is using two versions of the same browser to do regression testing. Using \tool in this setting actually removes a lot of the challenges we came across compared to the cross-browser setting. When using \tool for regression testing for Firefox, there are fewer false positives and negatives. The screenshots are largely consistent in shape and size when captured from two different versions of the same browser. Additionally, the regression test encounters fewer blocking issues, allowing an extra 180 screenshot sets (998 for the regression test vs 818 for the cross browser test) to be compared.

\begin{figure*}[t!]
\centering
\subfloat[Firefox screenshot]{\includegraphics[width=.9\columnwidth]{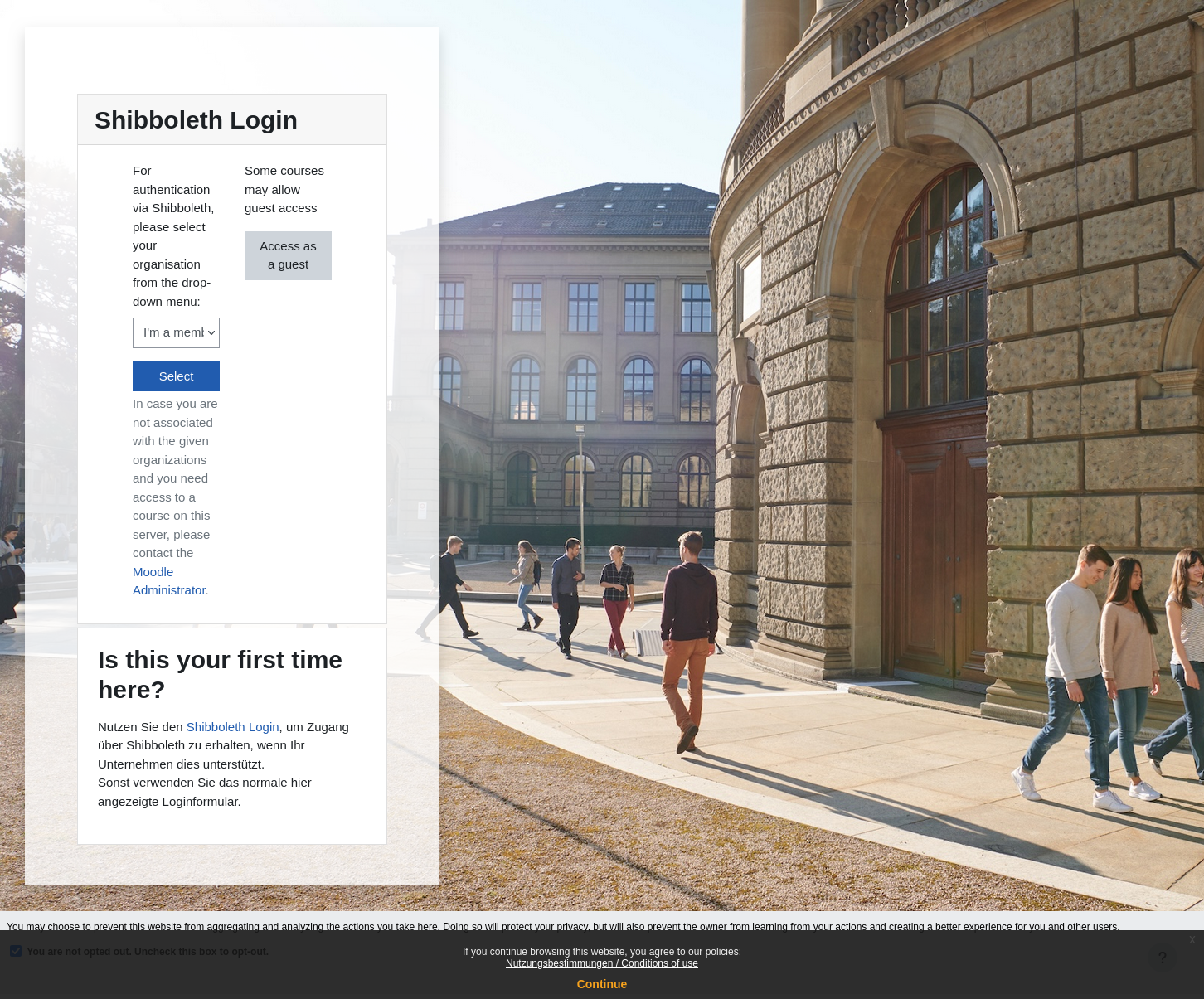}\label{fig:ff}}
  \hfill
  \subfloat[Chrome screenshot]{\includegraphics[width=.9\columnwidth]{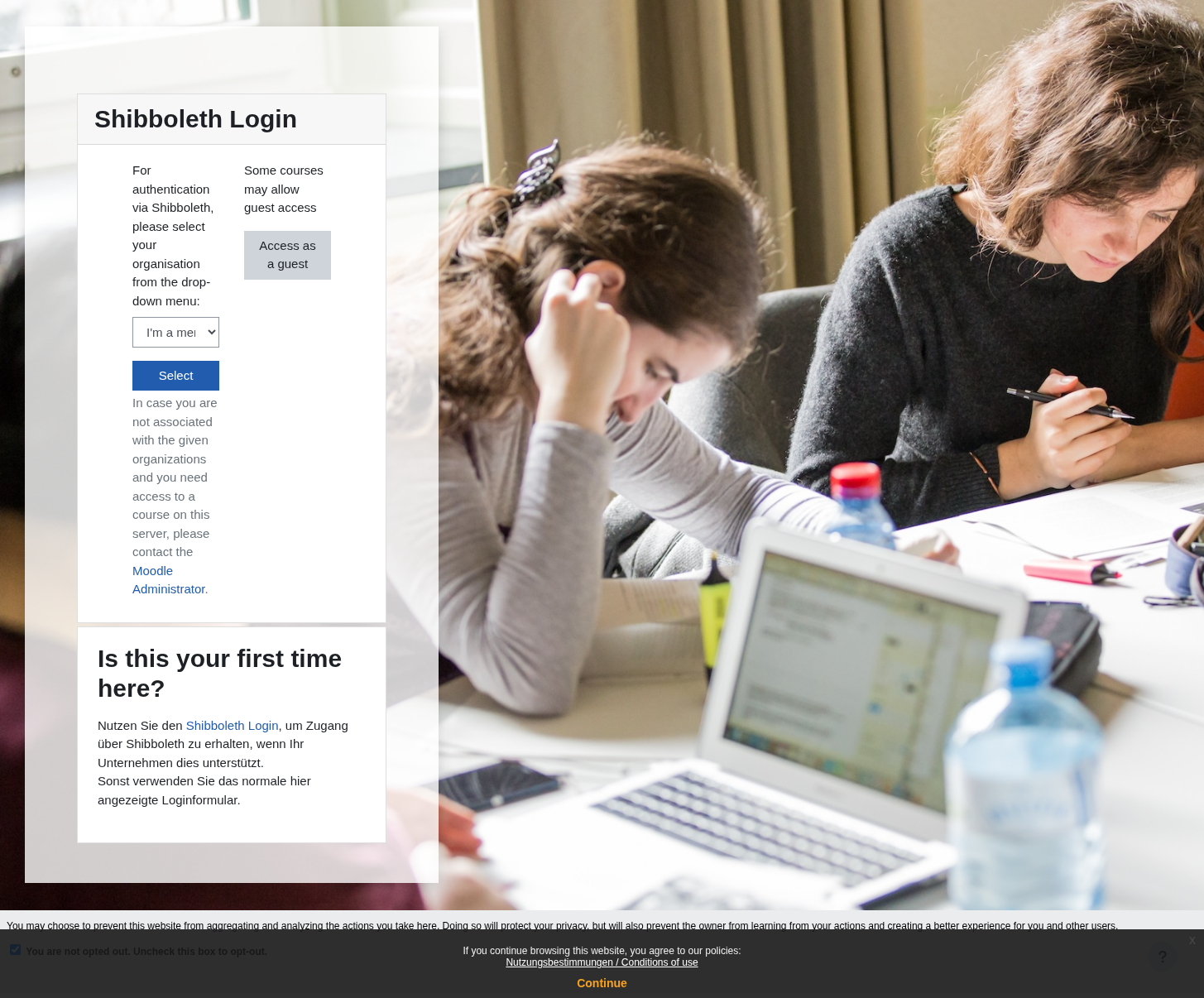}\label{fig:ch}}
\caption{Example of two screenshots taken of \url{https://moodle-app2.let.ethz.ch} with a dynamically changing background in Firefox and Chrome}
\label{fig:dyn_fp_ex}
\end{figure*}

\subsection{Lesson 3: Unaddressed pop-ups, such as cookie consent dialogs, can lead to false positives and should be explicitly handled} 
Pop-ups (or more precisely, modal dialogs), can vary in position, content, or behaviour across browser sessions. These pop-ups, such as cookie consent banners, subscription prompts, or advertisement overlays, can differ depending on timing, or whether the pop-up successfully loads. In some cases, a pop-up may appear in one rendering but not in another, leading \tool to falsely interpret this variation as an XBI.

To address false positives caused by pop-ups, we use Selenium to close the most common types of pop-ups. Selenium is provided with filters specifying the pop-up types it can attempt to close. However, the filters used to close pop-ups are limited in scope and cannot account for all variations, particularly less common or dynamically introduced pop-ups. As a result, some pop-ups persist and introduce visual differences that do not represent true XBIs. We found that the most effective strategy is to isolate results from \tool that contain information about pop-ups and place them into a separate table for review by developers. Thus, noise is reduced in the primary results and developers are able to focus on issues that are more likely to be genuine XBIs. We also preliminarily explored the use of web agents to automate the task of closing pop-ups. This approach appears promising, as the agents can identify and dismiss pop-ups without the need for predefined filters.

\subsection{Lesson 4: Some false positives are very difficult to prevent} 
These false positives are seen in scenarios involving dynamic elements such as changing pictures, backgrounds, or placeholder text. Because \tool processes a single static instance of the page, it often lacks sufficient context to recognize that these elements are dynamic. Although dynamic elements can create discrepancies between website screenshots, they are not considered XBIs, as such variations are expected on a website. For example, as seen in Figure~\ref{fig:dyn_fp_ex}, a login page with a different background image across visits might be flagged as an XBI, even though the variation is intentional and not a true inconsistency. The same issue occurs with placeholder text or images that change on each page reload. While one potential solution is to reload the site multiple times to capture variability, this approach is computationally expensive and time-consuming. 

In contrast, we noticed no instances where an advertisement is marked as a XBI. Ads, despite also being dynamic, often include indicators (such as labels or structural cues) that help \tool correctly identify them as non-critical changes, while dynamic elements may not have an indication of such. 

\section{Threats to Validity}
\label{Section:threats}

\textit{Construct validity:} Our approach intentionally ignores advertisements or dynamic elements to reduce false positives in XBI identification. While this reduces noise from expected content variability, it may also omit genuine XBIs in these elements. As a result, our approach currently misses XBIs that occur in dynamic elements or advertisements. This choice was intentional to reduce false positives in the generated reports. 

\textit{Internal validity:} A threat to the validity of the study stems from the use of Selenium to take the screenshots of the websites. Because an automation tool such as Selenium is used, some websites block the tool from taking screenshots, thus decreasing the number of websites analyzed. Future studies should consider looking into other web testing frameworks to take the screenshots, or settings for Selenium that would decrease the number of blocked screenshots. 

Changing VLM versions are also a potential threat to the validity of this study. New models are released at a rapid pace, making it challenging to keep evaluations up to date. As such, newer models may perform either better or worse than the ones used with \tool in this study. We designed \tool to be easily adaptable to different model versions, but future work should evaluate its performance with newer VLM releases.

A final threat to the internal validity of this study lies in the creation of the ground truth for evaluating \tool, which was manually verified by the first author. This process may introduce bias in labelling the impact of each bug report, a task that is already inherently ambiguous. To mitigate this risk, we verified a sample of the training data for fine-tuning with Mozilla developers.

\textit{External validity:} A potential threat to the validity of this study is that we only use bugs reported on Bugzilla or WebCompat. This website is managed by Mozilla employees and volunteers, thus, the web compatibility bugs may be biased toward issues related to Mozilla’s Firefox browser. Consequently, the dataset may not fully reflect the diversity of web compatibility bugs across all browsers.

Further, Selenium only supports a few browsers for taking screenshots, thus we only analyze Google Chrome and Mozilla Firefox. Though these browsers make up the majority of popular web browser engines, the results of this study may not generalize to compatibility bugs found exclusively on other bug reporting platforms or browsers. 
\section{Conclusion}
\label{Section:conclusion}

We introduce \tool, a tool to leverage vision language models to detect cross-browser inconsistencies by comparing website screenshots. We evaluate the effectiveness of both off-the-shelf (base and thinking) and fine-tuned VLMs with \tool in identifying XBIs across browser renderings. We find that both the base and fine-tuned versions of \tool perform well at this task, achieving an accuracy of 77\% and 79\%, respectively.

A cornerstone of our approach is to explicitly direct the VLM to identify advertisements and dynamic elements before analyzing for XBIs. To evaluate the foundation for this cornerstone, we show that VLMS are effective in identifying these elements, though the fine-tuned version shows a drop in accuracy for dynamic element detection. We also show that a fine-tuned base VLM can perform better than a thinking VLM at a fraction of the cost.

In a large-scale evaluation of 1,695 websites with no ground truth established, we found that reducing false positives, from ambiguous cases involving dynamic elements and pop-ups is essential for generating usable reports for developers. Furthermore, ensuring accurate XBI identification requires minimizing differences in how screenshots are captured across browsers.

Overall, this work demonstrates that \tool, especially when fine-tuned, is a approach for detecting of cross-browser inconsistencies in web development, helping to highlight discrepancies in website renderings that may indicate underlying bugs.

\bibliographystyle{IEEEtranS}
\bibliography{references}

\end{document}